\documentstyle{article}
\newcommand{\chapter}[1]{\centerline{{\bf{#1}\footnote
{Contribution to 
Festschrift volume for Englebert Sch\"ucking}}}} 
\newcommand{\chapterauthors}[1]{\centerline{{#1}}}
 \newcommand{\myl}[1]{\label{#1}\end{equation}}
\newcommand{\myref}[1]{(\ref{#1})}
\newcommand{\dA}{\fbox{\phantom{o}}}
\newcommand{\myla}[1]{\end{array}\myl{#1}}
\begin{document} 
\chapter{Gravity and the Tenacious Scalar Field}
\chapterauthors{ Carl H. Brans}
\begin{abstract}
Scalar fields have had a long and controversial life in gravity 
theories, having progressed through many deaths and 
resurrections. The first scientific gravity theory, Newton's, was
that of a scalar potential field, so it was natural for Einstein 
and others to consider the possibility of incorporating gravity 
into special relativity as a scalar theory.  This effort, though 
fruitless in its original intent, nevertheless was useful in     
leading the way to Einstein's general relativity, a purely 
two-tensor field theory.  However, a universally coupled scalar 
field again appeared, both in the context of Dirac's large number 
hypothesis and in five dimensional unified field theories as  
studied by Fierz, Jordan, and others.  While later 
experimentation seems to indicate that if such a scalar exists
its influence on solar system size interactions is negligible, 
other reincarnations have been proposed under the guise of 
dilatons in string theory and inflatons in cosmology.  This paper 
presents a brief overview of this history.
\end{abstract}
%\end{titlepage}
%\section{section one}\label{Sec:ch-1}
%\hd{Gravity and the Tenacious Scalar Field}
\section{Scalar Gravity?}
After  the conceptual foundations of special 
relativity had been laid by Einstein and the
natural  four-dimensional formalism for space-time had been
clarified by him, Minkowski and others, it was natural to 
consider how field theories should fit into the new framework.  
Of course, since it lay at the foundations of special relativity,
Maxwell's electromagnetic theory translated beautifully using the
 four-vector potential and two-form field formalism. The other 
classical field theory was gravity, so the question of 
incorporating gravity into the new relativity arose next. The 
standard textbook introductions to the subject naturally 
emphasize the logical path to Einstein's resolution of this, his 
general theory of relativity. Such pedagogical treatments  can 
even seduce the reader into believing that Einstein's general 
relativity is the logically necessary consequence of special 
relativity and the gravitational principle of equivalence.  The 
actual history was of course not so linear, and the logical path 
not so obvious to the participants.  The first, apparently most 
simple, approach is to generalize Newton's scalar 
gravitational theory to a special relativistic scalar one. In 
fact, this was precisely what was tried.  In the following we 
will sketch a brief overview of the physics of this period, 1907 
to 1915. This a is fascinating and highly instructive story. 
Fortunately, John Norton has provided an excellent, very 
readable,  review of the history of this subject, \cite{jn1}. 
\par Let us start with  the  natural four-force generalization of
Newtonian mechanics,
\begin{equation} {d\over 
d\tau}\Big(m{dx^\mu\over d\tau}\Big)={\cal F}^\mu,\myl{nm1}
 where the right side is the four-force vector.
Implicit in the physics of this equation is that the path 
parameter, $\tau$, must proper time, so the auxiliary condition 
\begin{equation}
 \eta_{\mu\nu}{dx^\mu\over d\tau}{dx^\nu\over 
d\tau}=-1,\myl{nm2}
 must be satisfied. 
 In modern terms this fixing of the 
path parameter would be described as a ``conformal gauge fixing 
condition.'' At any rate, \myref{nm1}, together with the 
assumption that $m$ is constant in the \myref{nm2} gauge, 
implies that
 \begin{equation} \eta_{\mu\nu}{\cal 
F}^\mu{dx^\nu\over d\tau}=0.\myl{nm3}
\par
As mentioned earlier,  Maxwell's electromagnetism 
 fits into special relativity quite naturally, since  
electromagnetism was the theory whose consistency triggered the 
re-investigation of space-time that ultimately led to it.  For 
electromagnetism the force in \myref{nm1} is
 \begin{equation} {\cal 
F}^\mu_{em}=F^{\mu\nu}\eta_{\nu\rho}{dx^\rho\over d\tau}, 
\myl{nm4} and \myref{nm3} follows neatly from the antisymmetry of
$F^{\mu\nu}.$ \par
Now, what of gravity?  In Newtonian-Galilean
 theory, the field can be 
described by a scalar potential, $\phi$, with
\begin{equation}
\nabla^2\phi={\kappa\over 2}\rho,\myl{ng1}
where $\rho$ is mass density, $\kappa\equiv 8\pi G,$ and $G$ the 
usual Newton constant. Using Galilean three-vector 
notation, \begin{equation}
{\bf E}_g=-\nabla\phi,\myl{ng2}
the equations of motion become
\begin{equation}
{d\over dt}\Big(m{d{\bf r}\over dt}\Big)=m{\bf E}_g.\myl{ng3}
Note that this potential has units of velocity squared, so that 
in the standard relativistic choice used in this paper, $c=1$, 
$\phi$ is dimensionless.
\par
 The natural, perhaps logically 
minimal, approach to including gravity in special relativity 
might thus seem to be ``relativizing'' \myref{ng1},\myref{ng2}, 
and \myref{ng3}, simply  extending them from three to four 
dimensions,
 \begin{equation}
\dA^2\phi={\kappa\over 2}\rho,\myl{sg1}
\begin{equation}
{\cal F}_g^\mu=-m\phi^{,\nu},\myl{sg2}
and \myref{nm1}. However,  \myref{nm3},
applied to \myref{sg2} results in
\begin{equation}
{dx^\mu\over d\tau}{\partial\phi\over\partial x^\mu}=
{d\phi\over d\tau}=0,\myl{sg3}
along the particle's path.  That is, the potential is constant
along the path of every particle, so the
gravitational force must necessarily be zero on every       
particle! Although the historical details are not entirely clear
 it seems likely that this is the problem to which 
Einstein referred in 1907 in discounting the appropriateness of a 
scalar special relativistic theory of gravitation without
allowing ``...the inert mass of a body to depend on the 
gravitational potential.''\cite{e1}  In the context of the time,
this seemed unacceptable, and so the problem of incorporating 
gravity into special relativity was the root of a great deal of 
concern on Einstein's part. It also provided fuel for criticism 
of the entire structure of special relativity by others, notably 
Abraham, \cite{jn1}. \par For our purposes, the next significant 
contribution was by  Nordstr\"om, 
\cite{gn}, who attacked the question of the possible dependence 
of mass on gravitational field directly, using, 
\begin{equation} 
m=m_0e^{\phi}.\myl{gn1} 
This is  a direct precursor to the idea that the 
time component of the metric, or perhaps the proper time itself 
(gauge dependence of mass)  depends on the gravitational 
potential. \par In  current 
formalism we guarantee consistency by starting from an action 
approach. For a single point particle of mass $m$, the field 
variable can be regarded as the path functions, $z^\mu(\tau)$, 
and the single particle action is \begin{equation} A_p=-\int 
m\sqrt{-\dot z^\mu\dot z_\mu} d\tau,\myl{sg4} where $\dot 
z^\mu\equiv dz^\mu/d\tau.$  To fit into the volume integration 
needed for field theories, this can be written \begin{equation} 
A_p=-\int \Big(\int m\sqrt{-\dot z^\mu\dot z_\mu} 
\delta^4(x^\mu-z^\mu(\tau))  d\tau\Big) d^4x.\myl{sg5}
 As a model for particle-field 
interaction consider electromagnetism,  with total field
 plus 
interaction actions,
\begin{equation}\begin{array}{ll}
A_{em}+A_I & =  {-1\over 16\pi}\int(A_{\mu,\nu}-A_{\nu,\mu})
(A^{\mu,\nu}-A^{\nu,\mu})d^4x+\\
& 
q\int\Big(\int\dot z^\mu(\tau)A_\mu
\delta^4(x^\nu-z^\nu(\tau)) d\tau\Big)d^4x.\myla{sg6}
Using $A_p+A_{em}+A_I$ as the total action results
in the Maxwell field equations with current
density source $J^\mu(x^\nu)=q\int\dot z^\mu(\tau)
\delta^4(x^\nu-z^\nu(\tau))d\tau,$ together with the correct
equation of motion for the particles.
Note that
the inertial mass, $m$, in $A_p$ is independent of the 
coupling constant, $q$ in $A_I$.\par
By analogy, the  total action for scalar gravity would be
\begin{equation}\begin{array}{ll}
A_p+A_I+A_\phi & =-\int \Big(\int m\sqrt{-\dot z^\mu\dot z_\mu}
\delta^4(x^\mu-z^\mu(\tau))  d\tau\Big) d^4x \\
& -
\int \phi \Big(\int m\sqrt{-\dot z^\mu\dot z_\mu}
\delta^4(x^\mu-z^\mu(\tau))  d\tau\Big) d^4x \\
&
-{1\over \kappa}\int\phi_{,\mu}\phi^{,\mu} d^4x.\myla{sg7}

Clearly the field variation results in \myref{sg1} with
$\rho(x^\mu)=m\int\delta^4(x^\mu-z^\mu(\tau))d\tau$ in the 
conformal gauge, \myref{nm2}.    However, the variation over
the particle's variables, $z^\mu(\tau),\dot z^\mu(\tau)$ results 
in something quite different from \myref{nm1} and \myref{sg2},
namely,
\begin{equation}
{d\over d\tau}\big(m(1+\phi)\dot z^\mu(\tau)\Big)=-m\phi^{,\mu}.
\myl{sg8}
This can of course be identified with Nordstr\"om's \myref{gn1},
with $\exp(\phi)\approx 1+\phi,$ to first order in $\phi.$
\par
While this provides a self-consistent and thus
potentially viable scalar gravitational theory, it was not an
attractive approach to Einstein, who seemed to have some strong
intuitive drive to incorporate the principle of equivalence at
the basis for the correct relativistic gravitational theory,
even though he apparently was not yet aware of the E\"otv\"os
experimental work on this matter.\par
 Einstein proceeded to criticize scalar 
special relativistic gravitational theories as being specifically
inconsistent with the equivalence principle.  When the 
four-vector equations of motion, say \myref{nm1} or \myref{sg8} 
are expressed in terms of local coordinate time, it is clear that
local coordinate acceleration of a particle will depend on the 
the particle's full kinetic energy, thus violating the equal 
acceleration principle.    For example, replacing the two
$d\tau$'s in the denominator of
the left side of \myref{sg8} with their coordinate expressions
gives 
\begin{equation}
{d^2{\bf r}\over dt^2}\sim -(1-v^2)\nabla\phi.\myl{ca1}
Thus, for example, spinning bodies would have smaller 
accelerations in a gravitational field than non-spinning 
identical ones, hot bodies than cold, etc.  Nordstr\"om pointed 
out that this effect would be too small to be measured by 
contemporary technology, but nevertheless it was regarded by 
Einstein as a serious obstacle to the consideration of such 
theories.  \par By a coincidence, von Laue was investigating the 
influence of special relativity on the theory of stress in 
bodies, and in so doing discovered what we now call the four 
dimensional stress-energy tensor, with $T^{00}$ identified with 
energy density, and $T^{ij}=p^{ij}$ the components of the spatial
stresses on the body, $p^{ij}=p\delta^{ij},$ for an isotropic 
fluid. Clearly a Lorentz velocity transformation would then mix 
the purely spatial stress components into the energy density, so 
that the energy density of a moving body would depend on its 
stress and quadratically on speed.  \par

At this point it is appropriate to take a close look at the 
concepts of active and passive gravitational mass, as well as 
inertial mass.  A complete study of these questions is beyond our 
scope here.  In fact, the extent to which Newton's laws are laws 
as opposed to simply definitions of force and mass is not an easy 
one to settle.  For a review of the ideas of Mach and others 
on these questions, see the book by Ray\cite{ray}.   Here we 
content ourselves with the following somewhat superficial review.
First assume  some adequate operational definition has been given
for Newtonian force.  Then gravitational force turn out to be 
proportional to a single scalar  parameter for a given particle 
at a particular spacetime point. This parameter is called ``{\it 
passive gravitational mass, $m_{gp}$}'', so 
\begin{equation}
{\bf F}=m_{gp} {\bf E_g}.\myl{defm1}
On the other hand, the gravitational field is determined by
field equations with source density proportional to ``{active
gravitational mass, $m_{ga}$}.''  It is now easy to see however, 
that these two parameters cannot be independent of other without 
violation of Newton's third law and thus conservation of 
momentum.  Thus, for a pair of particles, the force of particle 1 
on particle 2 is proportional to the $m^1_{ga}m^2_{gp}$ while the 
reaction force is proportional to  $m^2_{ga}m^1_{gp}$, so that
Newton's third law requires
\begin{equation}
{m_{ga}\over m_{gp}}=\mbox{ universal constant}.\myl{defm2}
Clearly this constant can be chosen to be one and  active and 
passive masses  assumed to be equal, unless Newton's third 
law is to be relaxed. The remaining mass is ``{\it inertial 
mass, $m_i$,}'' with
\begin{equation}
{\bf F}=m_i{\bf a}.\myl{defm3}
One form of the equivalence principle, the universality of 
gravitational acceleration at a fixed spacetime point, is then 
simply the statement that
\begin{equation}
{m_g\over m_i}=\mbox{ universal constant.}\myl{defm4}
\par
Returning to the historical matter of a scalar gravitational 
mass, the discovery by von Laue and others that special 
relativity would imply that stresses contribute to mass, the 
question was how to fit this into a gravitational theory.  
Nordstr\"om had already taken into account the contribution of 
the gravitational potential to mass, in \myref{gn1}, or in more 
modern formalism, \myref{sg8}, but still needed to account for 
the questions raised by von Laue's study of $T^{\mu\nu}.$ 
Specifically, what should be the $\rho$ on the right side of 
\myref{sg1}?  Nordstr\"om first considered $T^{00}$, evaluated 
in 
the rest frame of the particle, or invariantly $T^{\mu\nu}u_\mu 
u_\nu$, where $u_\mu$ are the components of the body's 
four-velocity.  Einstein suggested that the trace of the tensor, 
$T=T^\mu_\mu$, explicitly including stresses, would be more
appropriate.  However, what of stress energy tensors defined by 
null fields, such as electromagnetism, for which $T=0?$  Einstein 
pointed out that if such fields are to be treated as localized 
bodies, they must be contained and  their contribution to the 
gravitational field might be accounted for in terms of the 
stresses the confined fields exert on the container.  \par
At this point, the prospects for a scalar special relativistic 
gravitational theory looked good.   However, 
a critical thought experiment remained.  Consider a closed cycle
in which a stressed body (a rod)
 is lowered in a gravitational field, 
then unstressed and raised again.  Clearly energy is not 
conserved in this cycle.  The net gain in energy is associated 
with the lack of energy associated with a pure stress (no 
strain).  Nordstr\"om and Einstein were then able to show that 
energy conservation could be restored if they assumed that 
movement through the gravitational field was also associated with 
a change in length of the rod along the direction of the stress.
Thus work is done and energy released.  In other words, 
\par
{\bf Gravity and Geometry?}{\it A full working through of the 
implications of a special relativistic scalar gravitational theory 
leads to the suggestions that the lengths of rods, and 
rates of clocks, might depend on their location in a 
gravitational field.}\par
In modern notation Nordstr\"om's theory 
 would be expressed by saying that the 
metric is conformally flat,
\begin{equation}
ds^2=e^{2\phi}(dx^2+dy^2+dz^2-dt^2),\myl{ef1}
with the gravitational field, $\phi$ determined by
\begin{equation}
\mbox{`` $\dA$ ''}\phi={\kappa\over 2}T,\myl{ef2}
where the exact form of the source field equations 
was developed through several trial and error stages.
The crucial step of associating gravity with a distortion
of spacetime measurements, thus with empirical spacetime geometry 
had been taken.
Einstein and Fokker \cite{ef}  built on this work,
exploring the full 
geometric implications of Nordstr\"om's ideas.  Einstein noted   
that Nordstr\"om's geometric form, \myref{ef1}, conformally
flat, was too specialized and eventually
generalized to an arbitrary metric form, leading
to the full Einstein equations, with metric tensor playing the
role of gravitational potential. Apparently this  finally 
put an end to the search for a purely scalar special 
relativistic gravitational theory. Nevertheless, a universally 
coupled scalar field pops up in yet another context!
 \par 
\section{Kaluza-Klein theories} Very early in the development of 
relativity searches began for a unification of gravity with 
electromagnetism.  One of the most enduring of such attempts is 
that associated with the names of Kaluza and Klein.  The 
collection of papers on this subject edited by Appelquist et 
al\cite{ACF} provides a valuable resource for this subject, both 
classically and in more modern contexts. \par
Briefly, the Kaluza-Klein idea is
to embed the electromagnetic four-potential into the metric by 
enlarging the dimension of the space to five, and inserting the 
four-potential as \begin{equation} \gamma_{AB}=\pmatrix{V^2 & V^2
A_\beta \cr V^2 A_\alpha & g_{\alpha\beta}+V^2 A_\alpha 
A_{\beta}\cr}. \myl{kk1}
The vector tangent to the extra dimension is assumed to be a 
Killing vector, so all variables depend only on the spacetime 
coordinates, $x^\alpha.$ The existence of this Killing vector
gives rise to a natural foliation of the five-space of 
codimension 1, with local expressions as in \myref{kk1}.
Maintaining this foliation and
simply counting variables, we see the
usual 10 components of the spacetime metric, $g_{\alpha\beta},$ 
the four $A_\alpha,$ a four-vector, and a fifteenth, surprise 
quantity, a four-scalar, $V$. In the 1921 paper of Kaluza, 
\cite{k1}, this field is described as "noch ungedeutete,''
and left at that. In 1948 Thiry, \cite{ACF}, wrote out the full 
field equations five-Ricci, ${\cal R}_{AB},$ equals zero, 
explicitly, in effect taking this new scalar seriously as a 
possible universally coupled scalar field. The five-dimensional 
vacuum Einstein tensor  reduces to a scalar part,
 \begin{equation}
 {\cal S}_{55}={3V^2F_{\alpha\beta}F^{\alpha\beta}\over 
8}-{R\over 2},\myl{kk18s}
a four-vector set,
\begin{equation}
{\cal S}_{5\beta}= {3\over 2}V_{,\alpha}F^{\ \ 
\alpha}_{\beta}/2+ {V\over 2}F_{\beta\ \ \ ;\alpha}^{\ 
\ \alpha},\myl{kk19s}
and the four-tensor part,
\begin{equation}
{\cal S}_{\alpha\beta}=S_{\alpha\beta}+{V^2\over 
2}(F_{\alpha\mu}F^{\mu}_{\ \ \beta}+{\eta_{\alpha\beta}\over 
4}F_{\mu\nu}F^{\mu\nu})-({V_{,\alpha;\beta}\over 
V}-{\eta_{\alpha\beta}\dA^2 V\over V}).\myl{kk20s}
Here
\begin{equation}
dA={1\over 2}(A_{\alpha|\beta}-
A_{\beta|\alpha})\sigma^\beta\wedge\sigma^\alpha=
{1\over 2}F_{\beta\alpha}
\sigma^\beta\wedge\sigma^\alpha.\myl{kk15}
The full five dimensional analog of the Einstein equations are
\begin{equation}
{\cal S}_{AB}=0,\myl{kk30}
 derived from
\begin{equation}
\delta\int d^5x{\cal R}\sqrt{\vert g^{(5)}\vert}.\myl{kk31}
  However, the resulting 
\myref{kk18s},\myref{kk19s},and \myref{kk20s} contain the new 
field, $V$, apparently unrelated to either gravity or 
electromagnetism.  With the {\it ad hoc} choice,
\begin{equation}
V^2={\kappa\over 2\pi}=const,\myl{kk32}
the result is standard vacuum Einstein-Maxwell, 
\begin{equation}
F_{[\alpha\beta,\mu]}=0,\myl{kk33}
from \myref{kk15},
\begin{equation}
F_{\beta\ \ \ \ ;\alpha}^{\ \ \alpha}=0,\myl{kk34}
from \myref{kk19s}, and
\begin{equation}
S_{\alpha\beta}={\kappa\over 4\pi}(F_{\alpha\mu}F^{\ \ 
\mu}_\beta-{g_{\alpha\beta}\over 
4}F_{\mu\nu}F^{\mu\nu}),\myl{kk35}
from \myref{kk20s} plus one additional condition,
\begin{equation}
F_{\mu\nu}F^{\mu\nu}=0,\myl{kk36}
which follows from  ${\cal S}_{55}=0.$ Of course, this last 
condition is not a part of Maxwell's theory. To eliminate it
and reproduce the full Maxwell theory, the variational 
principle can be replaced by
\begin{equation}
\delta\int d^5x\sqrt{\vert g^{(5)}\vert}({\cal 
R}+\lambda(g_{55}-{\kappa\over 2\pi}))=0.\myl{kk37}
From this the combined vacuum Maxwell-Einstein field equations 
are recovered. However, there are some residual difficulties in 
the interpretation of the five-geodesic as the path of a particle
subject to both gravity and electromagnetic forces, so this 
approach has never been widely accepted as fully satisfactory in 
its original intent. Nevertheless, the basic ideas of the
Kaluza-Klein approach continue to be used in the wider sense 
of modern gauge theories, \cite{ACF}.  \par
From the viewpoint of this paper, however, Kaluza-Klein is 
important for introducing the new scalar $V$.  In
fact, from the identification required in \myref{kk32}, this 
new scalar can be associated with the Newtonian gravitational 
constant. \par
 \section{Dirac's Numbers} Meanwhile 
Dirac\cite{LNH}, building on the work of Eddington 
 and Milne, pointed out some remarkable 
clustering of dimensionless numbers composed from observed values
of certain fundamental constants. Dirac's starting point was the 
value of present cosmological age of the universe, $T_u$, as 
defined by the best available value of the Hubble constant in 
1938.  Of course, the  value of this number depends on 
arbitrary choice of units and so, of itself, cannot have any 
particular physical significance. Thus, another natural time unit
is needed. A natural choice is provided by any of a number of 
``atomic'' time scales, such as $e^2/m,$ or $\hbar/m$, where $e$ 
is the electronic charge and $m$ is some natural mass, such as 
that of the electron, or of a nucleon. Clearly, the range of 
numbers available for this choice is $10^3$, and this 
arbitrariness will not affect the agreement to orders of 
magnitude that follow. Dirac then noticed that this fundamental 
time ratio results in \begin{equation} t\equiv{T_u\over T_a} 
\approx 10^{40},\myl{drc1} where $T_a$ is one of the natural 
atomic time units.    Next Dirac decided to look at the 
dimensionless ratio of two of the fundamental forces, electrical 
and gravitational, on some standard atomic particle, 
\begin{equation} \gamma\equiv{e^2\over \kappa m^2} \approx 
10^{40}.\myl{drc2} Another number to consider is the ratio of the
present observed mass of the universe to the standard atomic 
mass, \begin{equation}
\mu\equiv{M_u/m}\approx 10^{80}.\myl{drc3}
These empirical numbers were first discussed by
Eddington, and
are generally known as Eddington numbers. From Dirac's
perspective,
the clustering of these natural,
dimensionless constants into groups of widely varying
magnitude $10^{40},10^{80}$ is remarkable indeed and led Dirac to
speculate that this clustering might well have some causal basis
in
some physical theory.  He proceeded to develop a cosmological
model in
which
\begin{equation}
\mu\approx t^2,\myl{drc4}
\begin{equation}
\gamma\approx t,\myl{drc5}
so that the quantities $\mu,\gamma$ change with the age of the
universe, the ``epoch'' in Dirac's terms.
Dirac's cosmology was later reconsidered in more detail 
by Canuto and others\cite{Can}.  At this point, Dirac's 
clustering leads to a number
\begin{equation}
{\mu\over t\gamma}\approx 10^0,\myl{drc6}
or, 
\begin{equation}
{\kappa M_u\over R}\approx 10^0.\myl{drc7}
\par Later, as a preface to inflationary cosmological 
models, these ``large number coincidences'' will play an 
important role. This will be discussed in the section on 
``inflatons,'' below.  However, at this point they raise the 
question of whether $\kappa$ is a truly universal constant, or 
would change in circumstances with different values for $M/R.$ In
other words, \myref{drc7} raises the possibility that $\kappa$ 
is  determined by the mass distribution in the 
universe.
\par \section{Scalar-Tensor Theories} \par Pascual 
Jordan \cite{SuW} was intrigued by the occurrence of the new 
scalar field in the Kaluza-Klein type of theories, and especially
its possible role as a generalized gravitational constant in the 
spirit of Dirac's hypothesis, \myref{drc7}. Building on this idea
Jordan and his colleagues, including Englebert Sch\"ucking, began
an investigation of Kaluza-Klein theories with special concern 
for the idea that the new five-dimensional metric component, a 
spacetime scalar, might play the role of a varying gravitational 
``constant.''  The resulting four-dimensional form of the field 
equations can be so interpreted.  However, Jordan and his 
colleagues took the next step of separating the scalar field from
the original five-dimensional metric context unified 
gravitational-electromagnetic context. Later Brans and Dicke 
\cite{bd} independently arrived at a similar proposal. Brans and 
Dicke were especially motivated by Mach's ideas on inertial 
induction.  Sciama\cite{Sci} had proposed a model theory of 
inertial induction, that is, a theoretical mechanism for 
generating the inertial forces felt during acceleration of a 
reference frame.  These forces were hypothesized to be of 
gravitational origin, occurring only during acceleration relative
to the ``fixed stars.''  In this model the ratio of inertial to 
gravitational mass will depend on the average distribution of 
mass in the universe, in effect making $\kappa$ a function 
of the mass distribution in the universe.\par  In 
commonly used notation  such theories  introduce
a scalar field, $\phi$, which will (locally and approximately) 
play the role of reciprocal Newtonian gravitational constant, 
$\kappa$. One obvious motivation for this choice of field 
quantity (rather than $\kappa$ itself) is Dirac's large number 
hypothesis in the form 
\begin{equation} 1/\kappa\approx 
M/R.\myl{lnh1} 
From this, rather than \myref{drc7}, we see the 
possibility that $1/\kappa$ itself might be a field variable and 
satisfy a field equation with mass as a source, something like
\begin{equation}
\mbox{`` $\dA$ ''}{1\over\kappa}=\rho.\myl{st0} 
\par Of course, 
the usual Lagrangian for Einstein theory including matter has 
$\kappa$ directly multiplying the matter contributions.  In this 
form, changes would be made to the local behavior of matter, the 
local equations of motion, as a result of variations in $\phi.$ 
 Consequently, in order to incorporate a Mach's principle by 
way of a variable gravitational ``constant,'' some modifications 
must be made to standard general relativity. Start by writing the
standard Einstein action as \begin{equation} \delta\int 
d^4x\sqrt{-g}(R+\kappa L_m)=0,\myl{sto-1} where  $L_m$ is the 
``usual'' matter Lagrangian, presumably derived from some 
classical or quantum model. 
\par
At this point it is useful to consider several aspects of the 
famous ``principle of equivalence.'' First is the statement 
that all bodies at the same spacetime point in a given 
gravitational field will undergo the same acceleration.  We 
will refer to this as the ``weak'' equivalence principle, WEP. 
 A stronger statement, on which standard Einstein's general 
relativity is built, is that the {\it only} influence of gravity 
is through the metric, and can thus (apart from tidal effects) be
locally, approximately transformed away, by going to an 
appropriately accelerated reference frame.  This is the 
``strong'' principle, SEP. If we start from an action of the 
form in \myref{sto-1} with variable $\kappa,$ we will be 
risking the geodesic equation for test particles, thus, 
possibly the WEP, and even mass conservation.  However, we are 
allowing for a possible violation of the SEP, since gravity, the 
universal interaction of mass, will influence local physics by 
changing the local $\kappa$.  As Dicke noted, the 
E\"otv\"os experiment verifies only the WEP and not the SEP, 
so, in the 1960's, it was reasonable to consider such 
alternatives.\par
Returning to the form of the action, let us then 
 isolate $\kappa$ from matter in the 
original \myref{sto-1} by dividing by it, \begin{equation} 
\delta\int d^4x\sqrt{-g}(\phi R+ L_m)=0,\myl{sto-2} where 
$\phi\equiv 1/\kappa.$    While we seem to have thus saved the 
geodesic equations for test particles, it is now known, of 
course, that the motion of composite bodies is more complex. It 
turns out that with refined observation techniques, even the 
coupling of $\phi$ directly to the gravitational field gives rise
to observable effects for matter configurations to which 
gravitational energy contributes significantly. This is now known
as the``Dicke-Nordtvedt'' effect and has been investigated 
in the earth-moon system with the lunar laser reflector.\par 
Nevertheless, let us proceed to see what follows  from \myref{sto-1}. We are anticipating 
field equations for $\phi$ so some action for this new field must
be supplied, \begin{equation} \delta\int d^4x\sqrt{-g}(\phi R+ 
L_m+L_\phi)=0.\myl{sto-3} The usual requirement that the field 
equations be second order leads to \begin{equation} 
L_\phi=L(\phi,\phi_{,\mu}).\myl{sto-4} Apart from this, there 
seem to be few {\it a priori} restrictions on $L_\phi.$  At first
glance, the standard choice for a scalar field, \begin{equation} 
L_\phi=-\omega\phi_{,\mu}\phi_{,\nu}g^{\mu\nu},\myl{sto-5} 
leading to a wave equation for $\phi$ with $R$ as source would 
seem to be natural.  However, the coupling constant $\omega$ 
would itself then need to have the same dimensions as the 
gravitational $\kappa$ that the new field is to replace! It would
at least seem reasonable to require that any new coupling 
constant be dimensionless for various reasons, so a natural 
minimal choice is \begin{equation}
L_\phi=-\omega\phi_{,\mu}\phi_{,\nu}g^{\mu\nu}/\phi,
\myl{sto-6}
in which $\phi$ has dimensions of inverse gravitational constant,
\begin{equation}
[\phi]=[\kappa^{-1}].\myl{sto-6.1}
The form \myref{sto-6} leads to an action which is often 
referred to as the ``Jordan-Brans-Dicke,''  JBD, action,
\begin{equation}
\delta\int d^4x\sqrt{-g}(\phi R+ L_m
-{\omega\over\phi}\phi_{,\mu}\phi_{,\nu}g^{\mu\nu})=0.
\myl{bd-1}
The variational principle, with standard topological and surface 
term assumptions, results in
\begin{equation}
\delta_m\int dx^4\sqrt{-g} L_m=0,\myl{bd-2}
\begin{equation}
\phi S_{\alpha\beta}= T_{(m)\alpha\beta}+\phi_{;\alpha;\beta}
-g_{\alpha\beta}\dA\phi+{\omega\over\phi}
(\phi_{,\alpha}\phi_{,\beta}-{1\over 2}g_{\alpha\beta}
\phi_{,\lambda}\phi^{,\lambda}),\myl{bd-3}
\begin{equation}
\omega({2\dA\phi\over\phi}-{\phi_{,\lambda}\phi^{,\lambda}
\over\phi^2})=-R.\myl{bd-4}
The first of these, \myref{bd-2}, is the standard variational 
principle for matter, which  follows the same equations as in 
Einstein theory, thus (apparently) satisfying the weak 
equivalence principle.  For test particles, \myref{bd-2}, results 
in the geodesic equations.  However, for extended, or composite,
 particles, this is may no longer be true, even in standard 
general relativity.  The second order interaction of matter by 
way of the scalar-metric coupling gives rise to  violations of 
the weak equivalence principle, so that bodies of different mass 
may have different gravitational accelerations in identical 
gravitational fields. Of course, because of the 
free standing $L_m$ in \myref{bd-1}, the energy tensor for matter
is still conserved, \begin{equation} T_{(m)\alpha\ 
;\beta}^{\phantom{(m)\alpha}\beta}=0. \myl{bd-4.1}
Taking the trace of \myref{bd-3}, solving for $R$,       
leads to another form for \myref{bd-4},
\begin{equation}
\dA\phi={1\over (2\omega+3)}T_{(m)},\myl{bd-6}
in which $T_{(m)}$ is the trace of the ordinary matter tensor. 
In a weak field model situation, within a static spherical 
shell of mass $M$, radius $R$ and otherwise empty universe this 
equation produces 
\begin{equation} \phi\approx \phi_\infty+{1\over 
4\pi(2\omega+3)}{M\over R}.\myl{bd-6.2}
If $\phi$ can be identified with the local reciprocal 
gravitational constant, and $\phi_\infty$ is set zero as a 
default asymptotic condition, then this equation is seen to be 
consistent with the Dirac coincidence, \myref{lnh1}.  Another 
natural approximation to \myref{bd-6} is to consider the effect 
of local matter over some background $\phi_0$ equal to the 
present observed value,
\begin{equation}
\phi\approx \phi_0+{1\over 4\pi}\sum_{\mbox{local m}}{m\over 
r}.\myl{bd-6.3}
 \par In equation \myref{bd-3} 
$T_{(m)\alpha\beta}$ are the components of the stress-energy 
tensor for matter derived from the matter Lagrangian $L_m$ in the
standard fashion.  This equation, which describes the sources of 
the gravitational field, can be re-written \begin{equation} 
S_{\alpha\beta}=(1/\phi)(T_{(m)\alpha\beta}+ 
T_{(\phi)\alpha\beta}).\myl{bd-5} This form clearly suggests that
$(1/\phi)$ does indeed act as a generalized gravitational 
``constant'', with both ordinary matter and the field $\phi$ 
itself serving as sources for the metric. However, it turns out 
that the presence of the $\dA\phi$ term on the right hand side of
\myref{bd-3}, together with \myref{bd-6} results in two 
occurrences of the matter tensor as a source, effectively 
producing a constant renormalization of $\phi$ as ``gravitational
constant.'' \par
The earliest serious investigations of this theory were by Jordan 
and his group, prominent among whom was Englebert Sch\"ucking.  
Heckman gave the first non-trivial exact vacuum solution, the
generalization of the Schwarzschild solution of standard Einstein 
theory.  Later, Sch\"ucking\cite{es} investigated the natural 
question of a Birkhoff type theorem for scalar-tensor equations. 
He was able to show that again the most general spherically 
symmetric solution must be static, if the scalar field is assumed
to be static, or to have a light-like gradient.  However, if 
$\phi$ is allowed to be a function of time, more general 
solutions can exist, of course.  A class of such solutions was 
also presented in Sch\"ucking's paper, and opened the way for 
studies of spherically symmetric, non-static, phenomena occurring
in scalar-tensor but not standard Einstein theory.\par
Early on questions of the choice of ``conformal gauge'' for the 
metric were considered.  In other words, replacing the metric, 
$g_{\mu\nu}\rightarrow \bar{g}_{\mu\nu}=\psi g_{\mu\nu}$
leads to a replacement of the action \myref{bd-1}
discarding the surface (topological) part, by
\begin{equation}
\delta\int d^4x\sqrt{-\bar{g}}({\phi\over\psi} 
\bar{R}+{3\phi\over 
2}{\vert\bar{\nabla}\psi\vert^2\over\psi^3}
-3\bar{\nabla}\psi\cdot\bar{\nabla}\phi/\psi^2+ L_m/\psi^2 
-{\omega\over\phi\psi}\vert\bar{\nabla}\phi\vert^2)=0. 
\myl{bd-1y} 
In particular, if $\psi$ is chosen to be $\phi$, 
\myref{bd-1y}becomes
\begin{equation}
\delta\int d^4x\sqrt{-\bar{g}}(\bar{R}-(\omega+{3\over 
2})\vert\bar{\nabla\alpha}\vert^2+e^{-2\alpha}L_m(\bar{g}))=0,\myl{cf-1}
where $\phi=e^\alpha$.  This variational principle is of course 
just the Einstein one for a massless scalar field(dimensionless),
$\alpha$, but universally coupled to all other matter through 
the $e^{-2\alpha}$ factor.  Regarding conformal rescalings of the
metric as a ``gauge,'' \myref{cf-1} is an expression of the 
theory in the ``Einstein gauge,'' as opposed to the original 
\myref{bd-1}, the ``Jordan'' gauge.  However, it should be clear 
that there is more to the conformal scaling than merely the 
formal expression of the equations.  In fact, the universal 
coupling of $\alpha$ to all matter in \myref{cf-1} means that in 
this metric test particles will not follow geodesics, nor have 
conserved inertial mass, etc., in the Einstein gauge.  In effect,
the identification of the Einstein metric used in the formulation
\myref{cf-1} as the ``physical'' metric leads to significant and 
observable violations of mass conservation and the WEP.  
Nevertheless, the choice of various conformal gauges continues to
be studied.
\par
 The investigation of such scalar-tensor 
generalizations of Einstein theory was strongly influenced by the
work of Dicke. In fact, the 1960's and 1970's saw an explosion of
interest in relativity and gravitational theories prompted at 
least in part by the presence of theoretically viable 
alternatives to standard Einstein theory, and Dicke's energetic 
promotion of them. By fortuitous coincidence this was also the 
time when NASA was coming of age and searching for space related 
experiments of fundamental importance.  Simultaneously, 
Nordtvedt, Will and others \cite{will} were led to provide 
rigorous underpinnings to the operational significance of various
theories, especially in solar system context, developing the 
parameterized post Newtonian (PPN) formalism as a theoretical 
standard for expressing the predictions of relativistic 
gravitational theories in terms which could be directly related 
to experimental observations. The equations of scalar-tensor 
theory approach those of standard Einstein theory as $\phi$ 
approaches a constant.  From \myref{bd-6.2} this would seem to 
occur in the limit of large $\omega$.  In fact, it is generally
true that the predictions of scalar-tensor approach those of 
Einstein for large $\omega,$ although there are interesting 
questions to be considered in general, \cite{dn}.
\par The ultimate outcome of 
these efforts was to set limits on the value of the parameter 
$\omega$ so large as to make the predictions of this theory 
essentially equivalent to those of standard Einstein theory.  In 
other words, solar system experimentation led to the conclusion 
that scalar-tensor modifications of standard Einstein theory 
would necessarily differ insignificantly from the standard, 
leading many workers to regard such theories as irrelevant. 
Nevertheless, as the next two sections show, universally coupled,
thus gravitational, scalar fields continue to play important 
roles in contemporary physics.\par 
\section{Dilatons}\par
 In the 
preceding discussion the scalar field was universally coupled 
to all matter and played a role  determining the locally 
measured Newtonian gravitational ``constant.'' Of course, scalar 
fields occur throughout physics, especially as quantum fields. 
Investigations of internal spaces for particle symmetries 
directly involve gauge theories with internal symmetry spaces 
occupied by families of fields which while having interesting 
transformation properties from the internal gauge group viewpoint
are nonetheless spacetime scalars. Some of the earliest  are 
 the $SO(N)$ bosons of the dual model, the 
Nambu-Goldstone bosons and the famous Higgs fields. Certainly, 
these scalars, as quantum fields, are based on different 
motivations than those leading to the scalar field in 
scalar-tensor theories. Nevertheless, the formalism, and perhaps 
macroscopic manifestations may turn out to be not too different. 
\par Historically, quantum dual models led to string theory and 
later superstring theory. In this process, a scalar field 
referred to as a ``dilaton'' appears quite naturally. This field 
couples to the trace of the two-dimensional string stress tensor.
It thus manifestly breaks the Weyl conformal (dilation) symmetry 
of the string. Nevertheless it is precisely what is needed to 
balance the quantum anomalies of this tensor by way of beta 
functionals of this tensor.  Along the way, the Einstein 
equations can be derived as the beta functions related to some 
external spacetime metric.  The two volumes of Green, Schwartz 
and Witten \cite{gsw} provide useful description of the origin 
and role of dilatons.\par
This is clearly a long and complicated subject, which we only
 summarize here. Consider a  string action as a 
natural generalization of a point particle action. For a background
metric, $g_{\alpha\beta},$ an obvious action choice is
\begin{equation}
S_1={-1\over 4\pi\alpha'}\int d^2\sigma \sqrt{\vert 
h\vert}h^{ab}\partial_aX^\alpha\partial_bX^\beta 
g_{\alpha\beta}(X^c),\myl{str1}
  with internal coordinate area 
$d^2\sigma,$  
internal string metric, $h_{ab},\ a,b...=1,2,$
and $\alpha'$  a tension related coupling parameter. If $S_1$ 
is compared to the relativistic point particle action, the role 
of point particle parameter has been replaced by the intrinsic 
surface metric, $h_{ab}.$  As in the point particle case, the 
derived physics should be independent of the internal 
parameterization, and in particular, the choice of string metric.
Thus, $\delta S_1/\delta h_{ab}=0,$ or 
\begin{equation} 
T_{ab}=\partial_aX\cdot\partial_b X=0.\myl{str2}
 Trivially, this implies 
that the trace of $T$ vanishes, but this would also be predicted 
by the conformal invariance of the string metric in \myref{str1}.
Actually, an even stronger result obtains: any two-dimensional 
metric is  conformally flat (but only locally, in general!),
 \begin{equation} 
h_{ab}=\phi\eta_{ab},\myl{str3}
 with constant $\eta_{ab}.$  Thus,
the surface element appearing in \myref{str1} reduces to    the 
flat one,
 \begin{equation}
  d^2\sigma\sqrt{\vert 
h\vert}h^{ab}=d^2\sigma\eta^{ab}.\myl{str4}
  The independence of 
the classical action from the choice of string metric is thus 
equivalent to invariance under Weyl (conformal) transformation 
internal to the string surface.  
In addition to $S_1$, it is natural to consider two additional 
terms, with associated fields derived from the string quantities.
 The first contains a spacetime two-form field, 
$B_{\alpha\beta},$ derived from the   
intrinsic volume two-form in the string,
\begin{equation}
S_2={-1\over 4\pi\alpha'}\int d^2\sigma 
\epsilon_{ab}\partial_aX^\alpha\partial_bX_\beta 
B_{\alpha\beta}(X^c).\myl{str3.x}
The second introduces the geometry of the string,
\begin{equation}
\chi={1\over 4\pi}\int d^2\sigma \sqrt{\vert 
h\vert}R^{(2)}.\myl{str4.x} 
However, one of the first discoveries 
relating geometry and topology was that this integral depends 
only on the topology of the string surface, and not the 
particular geometry.  This is in fact the first  Chern 
class for two dimensions.  The value for $\chi$ is the Euler 
number of the surface, and cannot be a dynamical variable.  
The two-geometry of the string can be introduced  non-trivially 
by adding to \myref{str4}  a scalar field, the 
``dilaton,'' $\Phi$, giving
\begin{equation}
S_3={1\over 4\pi}\int d^2\sigma\sqrt{\vert 
h\vert}\Phi(X^c)R^{(2)}.\myl{str5}
This term apparently breaks with the conformal invariance 
classically, thus violating  the desired invariance at the 
classical level.  However, paradoxically, it is precisely this 
term which can restore this invariance after quantization. Thus, 
when the action $S=S_1+S_2+S_3$ is quantized, conformal 
invariance is broken (an anomaly) unless the external fields 
satisfy three equations, as described in detail in GSW, volume 1,
page 180. For brevity, we drop the antisymmetric field, setting 
$B_{\alpha\beta}=0,$ and get (in the magical string dimension 
26!)  Einstein-like equation,
\begin{equation} 
0=R_{\alpha\beta}-2\Phi_{;\alpha;\beta},\myl{str6} 
\begin{equation} 0=4\Phi_{,\alpha}\Phi^{,\alpha}- 
4\Phi^{;\alpha}_{\ ;\alpha}+R.\myl{str7} Equivalently, these 
background field conditions can be derived from an ``effective 
action,'' \begin{equation}
\delta\int 
d^DXe^{-2\Phi}(R-4\Phi_{,\alpha}\Phi^{,\alpha}).\myl{str8}
  It is easy to verify then that this action is
 a special case of the vacuum scalar-tensor  one, \myref{bd-1},
 with $-2\Phi=\ln\phi,$ and $\omega=1.$  Nevertheless, it is 
difficult not to notice the close parallel between the 
universally coupled scalar of the old scalar-tensor theories 
and the new dilaton. \par
 \section{Inflatons} 
Cosmological models in standard general relativity have long been
known to contain serious conceptual difficulties. In particular, 
using standard general relativistic models, initial conditions 
must be fantastically fine-tuned in order to result in the 
universe as we now see it some $10^{10}$ years later.  See for 
example Peebles \cite{PJEP}, Linde \cite{al}. 
Consider the standard Robertson-Walker isotropic homogeneous 
metric model,
\begin{equation}
ds^2=-dt^2+R(t)^2d\sigma_\epsilon^2,\myl{cosm1}
where the three-space metric, $d\sigma_\epsilon^2$, is 
hyperbolic, flat, or spherical depending on whether $\epsilon$ 
is -1, 0 or +1. The Einstein equations result in 
\begin{equation}
\big({\dot R\over R}\big)^2=\kappa\rho/3+ {\epsilon\over 
R(t)^2}+\Lambda/3.\myl{dc1}
Defining the Hubble variable as usual, this can be rewritten,
\begin{equation}
1= \Omega+\epsilon\Omega_R+\Omega_\Lambda,\myl{cosm2}
where
 \begin{equation} 
 \Omega\equiv {\kappa\rho\over 
3H^2},\myl{dc2}
 \begin{equation}
  \Omega_R\equiv {1\over (R 
H)^2},\myl{dc3} 
and
 \begin{equation} \Omega_\Lambda\equiv 
{\Lambda\over 3H^2}.\myl{dc4}
Present data certainly gives  values for these three quantities 
each in the ballpark of one. In fact, \begin{equation}
\Omega(now)\approx {\kappa M\over R}\approx 10^0,\myl{cosm4}
which is one of Dirac's large number coincidences which was 
so instrumental in leading to the scalar-tensor theories.  
Now, however, we note it in the context of a universe evolving 
from earlier (``initial'') data drastically different from 
that at present.  For example, in the present matter 
dominated era the equation of state leads to
\begin{equation}
\rho R^3=M\approx const,\myl{cosm7}
whereas in an earlier radiation dominated state
\begin{equation}
\rho R^4\approx const.\myl{cosm8}
An analysis of the time evolution of these quantities under 
drastically different regimes show that an extremely small 
variation of the values of the $\Omega$'s at early times 
would result in drastically different values now.  But this is 
not the only conceptual problem.  For example, there are 
questions of how the universe could have homogenized itself from 
random early data (the ``horizon'' problem), and others, 
\cite{PJEP},\cite{al}\par
Guth\cite{G} pointed out that this myriad of difficulties could 
be at least partially resolved if the early stages of evolution 
were ``inflationary,'' that is
\begin{equation}
R(t)=R(0)e^{Ht},\myl{cosm9}
with constant $H.$  Such a model is consistent with \myref{dc1} 
for $\rho=\epsilon=0,\ \Lambda\ne 0.$  Of course, this is not 
consistent with present data, so something other than a 
cosmological constant is needed. One way to achieve it is to 
introduce a new massless scalar field was the 
``inflaton,'' $\phi,$ with Lagrangian density, \begin{equation} 
{\cal L}=g^{\alpha\beta}\phi_\alpha\phi_\beta-V(\phi).\myl{inf2} 
The resulting stress tensor produces an effective mass density 
and pressure given by \begin{equation} \rho_\phi=\dot\phi^2/2+V,\
\ p_\phi=\dot\phi^2/2-V.\myl{inf3}
By ``fine-tuning'' the potential, $V$, at least some, but 
certainly not all, of the problems discussed above can be 
resolved.  In some versions, the inflaton  has a dilaton-like 
nature, in others it is reminiscent of the $\phi$ in the old 
scalar-tensor theories, with $\omega$ so large as to make the 
deviations from general relativity insignificant in contemporary 
solar system physics, but very significant in earlier 
cosmological contexts.  At present, it seems likely that more 
than one scalar field will be required.
The entire field of inflationary models is a very active one at 
present with many competing models. However, the role of scalar 
fields such as $\phi,$ is prominent in many of them. \par
 \section{Conclusion}
  Universally coupled, thus 
gravitational, scalar fields are still active players in 
contemporary theoretical physics. So, what is the relationship 
between the scalar of scalar-tensor theories, the dilaton and the
inflaton? Clearly this is an unanswered and 
important question. The scalar field is still alive and 
active, if not always well, in current gravity research.
\par 
\section{Acknowledgments}
Bob Dicke led the way for me in the late 1950's, starting with 
inertial forces, Mach's principle all the way through the ideas 
of scalar-tensor modifications of Einstein's general relativity. 
His death in March of 1997 deprives all of us in relativity of a 
source of ideas, and perhaps more importantly, a driving spirit 
of enthusiasm for understanding the mysteries of gravity.\par
John Norton was of great help teaching me some of the interesting
early history of scalar gravity and its role in the development 
of General Relativity.  \par  \end{document}